\documentclass[preprint,12pt]{elsarticle}
\usepackage{amssymb}
\usepackage{url}
\usepackage{color,soul}
\usepackage{amsmath}
\usepackage{float}
\usepackage{hyperref}
\usepackage{makecell}
\newcommand\aug{\fboxsep=-\fboxrule\!\!\!\fbox{\strut}\!\!\!}

\journal{Computer Physics Communications}

\begin{document}

\begin{frontmatter}



\title{OstravaJ: a tool for calculating magnetic exchange interactions via DFT}


\author[inst1]{Jan Priessnitz}

\affiliation[inst1]{organization={IT4Innovations, VŠB - Technical University of Ostrava},%Department and Organization
            addressline={17. listopadu 2172/15},
            city={708 00 Ostrava-Poruba},
            % state={State One},
            country={Czech Republic}}

\author[inst1]{Dominik Legut}
% \author[inst1,inst2]{Author Three}

% \affiliation[inst2]{organization={Department Two},%Department and Organization
%             addressline={Address Two},
%             city={City Two},
%             postcode={22222},
%             state={State Two},
%             country={Country Two}}

\begin{abstract}
OstravaJ is a Python package for high-throughput calculation of exchange interaction terms in the Heisenberg model for magnetic materials. It uses the total energy difference method, where calculations are based on the total energy of the system in different magnetic configurations, calculated by means of density functional theory. OstravaJ can propose a suitable set of magnetic configurations, generate VASP configuration files in cooperation with the user, and read VASP calculation results, which minimizes necessary human interaction. It can also calculate other relevant properties (e. g. MFA and RPA critical temperature, spin-wave stiffness) and provide input for various atomistic spin dynamics codes.

We present results for a number of materials from various classes (metals, transition metal oxides), compared to other methods. They show that the total energy difference method is a useful method for exchange interaction calculation from first principles.

\end{abstract}




\end{frontmatter}


\section{Introduction}

Density functional theory (DFT) is a widespread method which can be used to calculate many properties of condensed materials from first principles. Gradual improvements in the accuracy of DFT have recently allowed us to focus on magnetism in materials -- usually operating on the meV-eV per atom energy scale.

To better understand magnetism in condensed matter, several microscopic models of magnetism were introduced. A very successful example is the classical Heisenberg model, which reduces the system into a set of rigid spins on atoms influenced by interactions of various origins. This model is simple enough to allow for simulating magnetic systems on greater scales (atomistic spin dynamics, micromagnetics), but still flexible enough to accurately describe the majority of magnetic materials. The most important interaction in the classical Heisenberg model is the exchange interaction, coupling pairs of spins.
The strength of the exchange interactions influences many properties of a magnetic material, deeming it suitable or unsuitable for applications. A computational method for calculating exchange interactions from first principles can significantly accelerate the search for new magnetic materials and the development of new electronic devices.

There are several methods for calculating exchange interactions. A popular method is based on the work of Liechtenstein, Katsnelson, Antropov, and Gubanov -- the LKAG formula, or the magnetic force theorem \cite{LIECHTENSTEIN198765,pajda_ab_2001}. Principally a linear-response method, it calculates exchange interactions by infinitesimally tilting one spin and calculating the resulting energy correction via perturbative theory and Green's function. A prerequisite for this approach is a DFT calculation of the ground state in a localized basis set. Results in non-local basis sets need to be expressed in terms of Wannier functions \cite{PhysRevB.91.224405}, for example, via the Wannier90 code \cite{wannier90_1,wannier90_2}. A popular code for automating this calculation is the TB2J package \cite{HE2021107938}. Time-dependent DFT can also be used \cite{GORNI2022108500}.

We present the OstravaJ package, which aims to automate an alternative, conceptually simpler method of calculating exchange interaction energies -- the total energy difference method \cite{https://doi.org/10.1002/jcc.25081,xiang_magnetic_2012}. It is based on calculating the total energy of several magnetic configurations (magnetic phases) using DFT and then mapping the total energies onto the Heisenberg Hamiltonian of the system. The result is a system of linear equations with exchange interaction energies as variables. A significant advantage of this method is that it does not assume any requirements on the basis set of the DFT calculation. On the other hand, some problems may arise when calculating the total energy of the excited magnetic configurations. Furthermore, finding suitable  magnetic configurations to calculate long-range interaction energies becomes increasingly difficult. OstravaJ introduces a novel scheme for suitable magnetic configuration selection which aims to tackle this challenge.

\section{Methods and implementation}

\subsection{Total energy difference method}

Let us have an infinite periodic system of magnetic ions with atomic magnetic moments pointing in the direction defined by the spin vectors $\vec{S_i}$. The system can be described by the following classical Heisenberg Hamiltonian:

$$\mathcal{H} = \mathcal{H}_0 - \sum_{<ij>} J'_{ij} \vec{S_i} \cdot \vec{S_j}~~~~~~\vec{S_i} \in \mathbb{R}^3, |\vec{S_i}| = 1$$

where $\mathcal{H}_0$ denotes the non-magnetic part of the energy and $J'_{ij}$ is the strength (energy) of the exchange interaction between spins $\vec{S}_i$ and $\vec{S}_j$. The summation goes over all spin pairs once.

Due to the spatial symmetry of the system, we can assume that certain exchange interactions must be equal. We can thus sort all spin pairs $(i, j)$ equivalence classes $X_k$ based on the exchange interaction energies $J'_{ij}$ and define $J_k$ -- a representative of $k$-th exchange interaction equivalence class, such that $\forall (i,j) \in X_k: J'_{ij} = J_k$.

The Heisenberg Hamiltonian of the system can then be expressed as

$$\mathcal{H}_{\mathrm{Heis}} = \sum_k J_k c_k$$
$$c_k = \sum_{<ij> \in X_k} \vec{S}_i \cdot \vec{S}_j$$

In an infinite system, the number of equivalence classes is also infinite. OstravaJ introduces a distance cutoff $d_{\mathrm{max}}$, where only spin pairs separated by distance smaller than $d_{\mathrm{max}}$ are considered. This limits the number of equivalence classes to $N_k$ and all other exchange interaction energies are considered zero. The Heisenberg Hamiltonian then becomes

$$\mathcal{H}_{\mathrm{Heis}} = \sum_{k < N_k} J_k c_k$$

Let us say that we have defined $N_l$ concrete magnetic configurations and calculated their total energies $E^{(l)}$. Let us denote the \textit{final} (converged) direction of $i$-th spin in $l$-th magnetic configuration as $\vec{S}_i^{(l)}$ and $k$-th coefficient as $c_k^{(l)}$

$$c_k^{(l)} = \sum_{<ij> \in X_k} \vec{S}_i^{(l)} \cdot \vec{S}_j^{(l)}$$

For each magnetic configuration, we can construct an expression for the energy $E^{(l)}$ with $J_k$ as free variables

$$E^{(l)} = E_0 + \sum_{k < N_k} J_k c_k^{(l)}$$

where $E_0$ is the non-magnetic part of the total energy.

We can now assemble a system of linear equations with $N_l$ equations and $N_k + 1$ free variables ($J_1$,...,$J_{N_k}$ and $E_0$) which can be represented by the following matrix for the case of $N_k = 3$ and $N_l = 4$:

\begin{equation}
\begin{pmatrix}
    c_1^{(1)} & c_2^{(1)} & c_3^{(1)} & 1 &\aug& E^{(1)} \\
    c_1^{(2)} & c_2^{(2)} & c_3^{(2)} & 1 &\aug& E^{(2)} \\
    c_1^{(3)} & c_2^{(3)} & c_3^{(3)} & 1 &\aug& E^{(3)} \\
    c_1^{(4)} & c_2^{(4)} & c_3^{(4)} & 1 &\aug& E^{(4)} \\
\end{pmatrix}
\end{equation}

Solving this matrix yields a vector of values $(J_1, J_2, J_3, E_0)$. The system of linear equations can then be solved exactly if $N_l = N_k + 1$ or by the least-squares method if $N_l > N_k + 1$. Note that a unique solution exists only if the matrix has full rank and the vectors represented by the matrix rows form a basis spanning the full $(N_k + 1)$-dimensional space. In the opposite case, a non-unique solution for $J_k$ carries little physical information. This requirement can be satisfied by choosing a suitable set of magnetic exchange interactions, which is described later in this article.

The solution of this system consists of the exchange interactions $J_1, ..., J_{N_k}$ and the non-magnetic part of total energy $E_0$. When $N_l > N_k + 1$, the error of the least-squares fitting is also available.

The non-magnetic total energy $E_0$ can be further used to easily calculate the critical temperature of the system in the mean-field approximation (MFA). Let $E_\mathrm{gs}$ be the total energy of the ground state and $N_s$ be the number of magnetic sites in the unit cell. The MFA critical temperature can then be calculated as:

\begin{equation}
    T_C^{MFA} = \frac{\Delta E}{3 k_b N_s} = \frac{E_0 - E_{gs}}{3 k_b N_s}
\end{equation}

%%\subsubsection{Example: fcc structure}

\subsection{Choosing a set of magnetic configurations}

First task within the total energy difference method is to find a suitable set of magnetic configurations which leads to a system of linear equations with a unique solution, as described in the section above. This task depends on the interaction distance cutoff and, in turn, the number of unique exchange interactions to take into account (denoted $N_k$) -- a parameter configured by the user beforehand. The configuration set must include at least $N_k + 1$ configurations which yield linearly independent Heisenberg Hamiltonians.

Searching for magnetic configurations within the primitive unit cell of the material in order to find the suitable set is usually unsuccessful. In the extreme case of a one-atom unit cell, only the ferromagnetic configuration is accessible, and the configuration set can never be assembled. Before the search, the unit cell must be extended to a supercell containing $N_x \times N_y \times N_z$ unit cells. The supercell must be large enough so that a suitable set of $N_k + 1$ independent configurations exists, but not too large so that the memory and computational requirements of the subsequent calculations are reasonable. As of now, the suitable interaction cutoff and supercell size need to be configured by the user. OstravaJ will be able to select the best supercell size automatically in the future.

For a unit cell with $N_s$ magnetic sites, the number of possible collinear magnetic configurations $N_c$ grows exponentially as $N_c = 2^{N_s}$. This makes brute-force traversal of the whole configuration space very impractical. The novelty of OstravaJ lies in the special algorithm for generating the suitable configuration set, able to search through space consisting of hundreds of magnetic sites and more than $2^{100}$ individual configurations.

OstravaJ builds the configuration set incrementally, adding configurations into the set one by one. A configuration can be added only if it fulfills the following criteria:
\begin{enumerate}
\item new configuration is linearly-independent to the current set of configurations -- if it were added, the rank of the resulting system of equations matrix would increase by 1.
\item equivalent magnetic sites must have identical local magnetic surrounding (must be chemically equivalent)
\end{enumerate}

Let us look more closely at the second requirement of the identical local magnetic surrounding. This can be expressed in terms of the Heisenberg Hamiltonian. First, let us divide the Hamiltonian into contributions made by individual magnetic sites, denoted as $H_i$

\begin{equation}
\mathcal{H} = \sum_{<ij>} J_{ij} \vec{S_i} \cdot \vec{S_j} = \frac{1}{2} \sum_{i} H_i
\end{equation}
\begin{equation}
H_i = \sum_{j} J_{ij} \vec{S_i} \cdot \vec{S_j} = \vec{S_i} ( \sum_{j} J_{ij} \vec{S_j})
\end{equation}

The second requirement says that individual contributions $H_i$ of all magnetic sites of equal type must be equal

\begin{equation}
\forall i,j: H_i = H_j
\end{equation}

\begin{figure}[H]
    \centering
    \includegraphics[width=0.5\linewidth]{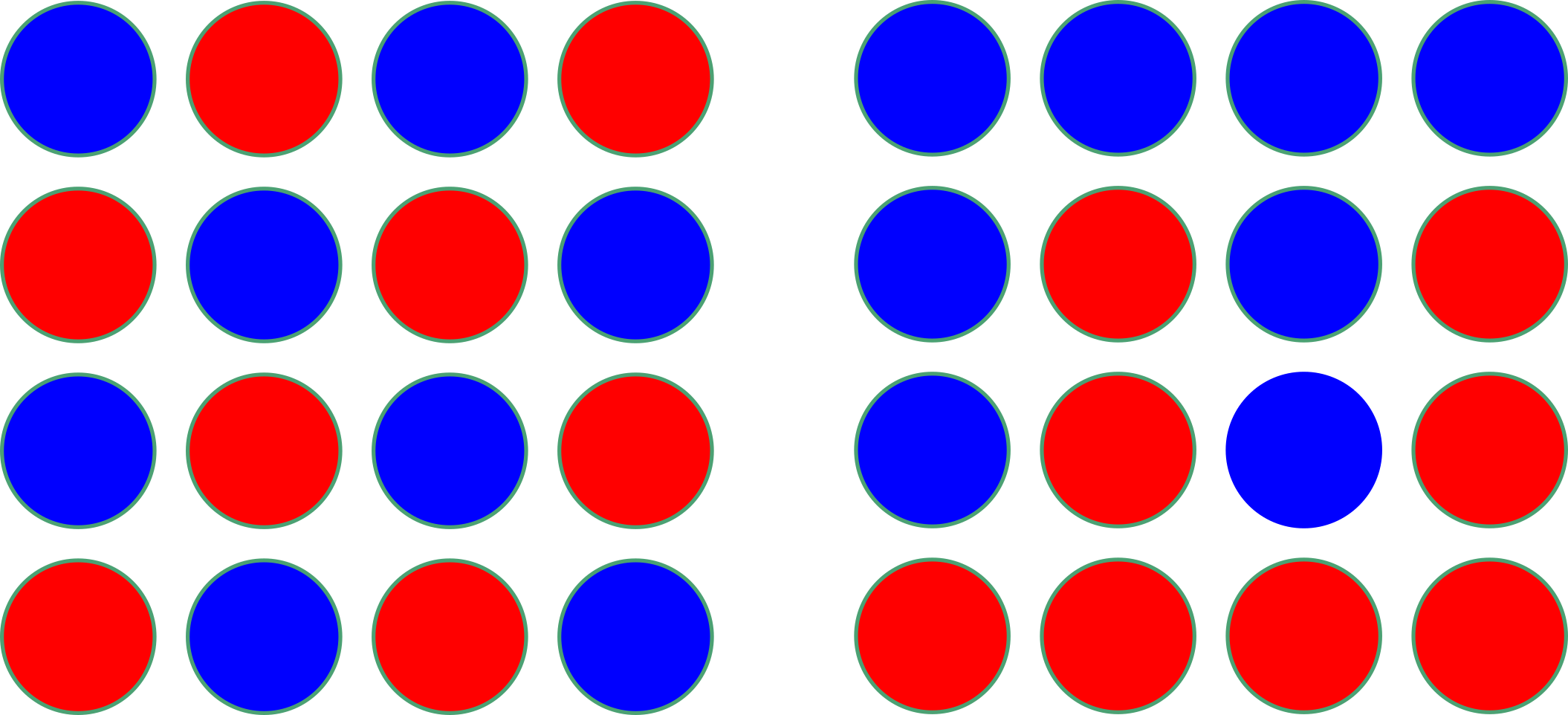}
    \caption{Example of the identical magnetic surrounding criterion. Blue circles show magnetic sites with spin up, red circles show spin down. On the left, the criterion is fulfilled. On the right, the criterion is broken -- various sites have various surroundings.}
    \label{fig:identical-surrounding-example}
\end{figure}

\autoref{fig:identical-surrounding-example} shows an example of the identical magnetic surrounding criterion on a 2D square lattice. The configuration on the left fulfills the criterion -- all sites have 4 nearest neighbors pointing antiparallel, 4 next-nearest neighbors pointing parallel, etc. The configuration on the right does not fulfill the criterion -- there are sites with 1 nearest neighbor parallel and 3 antiparallel, sites with 2 parallel and 2 antiparallel, and sites with 3 parallel and 1 antiparallel. Various magnetic sites have different local magnetic surroundings.

\begin{figure}
    \centering
    \includegraphics[width=\linewidth]{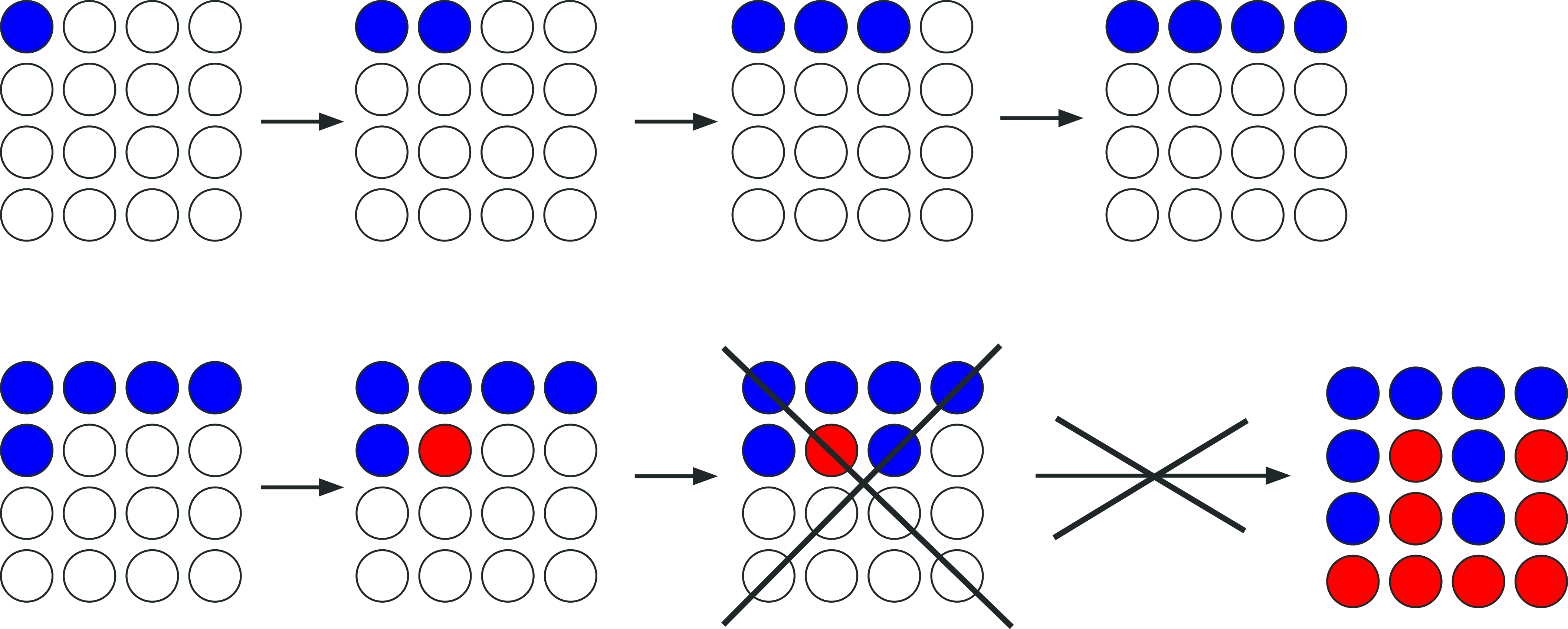}
    \caption{Example of the recursive traversal and early abortion. In the 7th step, the algorithm is already certain that there is no way to fill the rest of the spins and fulfill the identical magnetic surrounding criterion and it aborts that branch of the search.}
    \label{fig:early-abort}
\end{figure}

The performance in searching through the configuration space achieved by OstravaJ lies in the fact that it traverses the space recursively and checks whether the identical magnetic surrounding criterion is reachable even in incomplete configurations. In case it is not reachable, it aborts that branch of the traversal early and saves a significant amount of steps. An example of recursive traversal and early abortion is shown in \autoref{fig:early-abort}.

%\subsection{Running DFT calculation}

%<mention VASP>
%<mention pymatgen>

%\subsection{OstravaJ user interface}

%<<CLI>>

%<<as pymatgen library>>

%\section{Implementation}

%\section{Usage}

% This section describes regular usage of current version of the OstravaJ package. Instructions are also available in the \texttt{README.md} file in the source code.

\subsection{Installation}

OstravaJ is available at \url{https://code.it4i.cz/jpriessnitz/ostravaj} as a Git repository. The package itself does not require explicit installation, but it depends on several Python packages, as listed in the \texttt{requirements.txt} file. Dependencies can be installed via

\begin{verbatim}
pip3 install -r requirements.txt
\end{verbatim}

OstravaJ can then be used through
\begin{verbatim}
<path-to-OstravaJ>/OstravaJ.sh <subcommand> <args>
\end{verbatim}

Apart from OstravaJ, users must also install a DFT software (currently, only VASP can be used) to conduct complete calculations. However, OstravaJ does not invoke the DFT software directly, so it does not have to be configured for cooperation.

Further instructions on how to use OstravaJ are available in the \texttt{README.md} file in the source code.

\subsection{Workflow overview}

% Complete calculation of exchange interactions for a particular material involves several steps, shown in \hl{Fig. 69}.

The OstravaJ tool is a Python package which aims to automate the total energy difference method for calculating exchange interactions. It requires external DFT software to conduct the total energy calculations. Currently, the only supported DFT code is VASP \cite{VASP1,VASP2}, but we plan to implement support for other popular DFT codes in the future.

Let us now look at the whole calculation workflow from the user perspective. We will look at each step in more detail in later sections of this article.
% Fig. 69 shows an overview of the workflow.
% <<figure>>

Before the calculation begins, user must provide several things:

\begin{itemize}
    \item crystal structure in the form of a \texttt{POSCAR} file and information about which ions are to be considered magnetic
    \item size of the magnetic unit cell
    \item distance cutoff or a selection of exchange interactions
    \item base spin
    \item configuration files for the VASP DFT calculation (\texttt{INCAR}, \texttt{KPOINTS}, \texttt{POTCAR}, ...)
\end{itemize}

OstravaJ then begins to search for a suitable set of magnetic configurations based on the crystal structure, magnetic unit cell size and distance cutoff. Either it succeeds with the search and outputs a number of VASP configurations, or it fails to find a suitable set which would lead to solving a system with the desired set of exchange interactions. The number of suitable configurations depends on the magnetic unit cell size -- more complicated magnetic configurations are accessible in larger magnetic unit cells. In general, the larger the number of exchange interactions, the larger the magnetic unit cell must be.

As a next step, the user must launch the VASP calculations generated by OstravaJ.

After converging all DFT calculations, OstravaJ reads the converged magnetic moments and total energies, assembles the system of linear equations and solves for exchange interaction energies. It also calculates some experimentally important quantities such as critical temperature in the mean-field approximation (MFA) and random-phase approximation (RPA) and spin-wave stiffness.

Optionally, OstravaJ can also prepare system configuration for atomistic spin dynamics simulations and calculate the magnetization vs. temperature curve. It automatically determines the spin-dynamics critical temperature which is generally considered more precise compared to MFA and RPA (usually $T_{\mathrm{C,RPA}} < T_{\mathrm{C,ASD}} < T_{\mathrm{C,MFA}}$. Currently, only UppASD code \cite{skubic_method_2008} is supported for spin dynamics calculations.

\section{Results}

We showcase the capabilities of OstravaJ on calculations of several simple structures. The results are compared to both computational and experimental data from other sources to see the accuracy of the total energy difference method. All DFT calculations were conducted with the Vienna Ab initio Simulation Package (VASP) software (version 6.4). VASP is a plane-wave basis set implementation \cite{VASP1,VASP2} of the Density Functional Theory within the Projector Augmented Wave (PAW) approximation \cite{PAW}.

\subsection{Transition metal oxides}

Firstly, we calculate magnetic exchange interactions in transition metal oxides CoO, NiO and MnO. These materials all have rock-salt structure which has face-centered cubic lattice and a unit cell containing one atom of oxygen and one atom of transition metal. They are dielectric materials, meaning that long-range exchange interactions are negligible.

As a first step, we provide the material structure files to OstravaJ and set the exchange interaction cutoff to 6 nearest exchange interactions, meaning that OstravaJ needs to find at least 7 unique (linearly independent) magnetic configurations. The supercell size was set to $4 \times 4 \times 2$ unit cells -- OstravaJ did not find enough independent configurations in smaller supercells.

Total energies of all configurations were then calculated using VASP with calculation parameters shown in table \autoref{tab:TMO-params}. We used the generalized-gradient-approximation with the Perdew-Burke-Ernzerhof (PBE) exchange-correlation functional \cite{PBE} with phenomenological Hubbard U via the Dudarev approach \cite{Dudarev} to capture the electron correlations. In addition, the nearest two exchange interactions in MnO and NiO were recalculated with the HSE06 hybrid exchange-correlation functionals \cite{heyd_hybrid_2003}.

\begin{table}[H]
\centering
\begin{tabular}{|c|c|c|c|c|c|}
     \hline
     & \makecell{MnO\\(PBE+U)} & \makecell{MnO\\(HSE06)} & \makecell{NiO\\(PBE+U)} & \makecell{NiO\\(HSE06)} & \makecell{CoO\\(PBE+U)} \\
     \hline
     lattice parameter [\AA] & 4.45 & 4.45 & 4.20 & 4.20 & 4.25 \\
     \hline
     energy cutoff [eV] & 700 & 500 & 500 & 500 & 700 \\
     \hline
     Hubbard $U - J$ [eV] & 5.0 & -- & 5.0 & -- & 4.0 \\
     \hline
     supercell & $4 \times 4 \times 2$ & $2 \times 2 \times 2$ & $4 \times 4 \times 2$ & $2 \times 2 \times 2$ & $4 \times 4 \times 2$ \\
     \hline
     k-grid & $2 \times 2 \times 4$ & $6 \times 6 \times 6$ & $2\times2 \times 4$ & $6 \times 6 \times 6$ & $2 \times 2 \times 4$\\
     \hline
     \makecell{energy convergence\\criterion [eV]} & $10^{-7}$ & $10^{-7}$ & $10^{-7}$ & $10^{-7}$ & $10^{-7}$\\
     \hline
     \makecell{no. of valence\\electrons (M+O)} & 7+6 & 13+6 & 16+6 & 16+6 & 17+6 \\
     \hline
\end{tabular}
\caption{Important parameters for the transition-metal-oxide DFT calculations.\label{tab:TMO-params}}
\end{table}

\begin{figure}[H]
    \centering
    \includegraphics[width=1\linewidth]{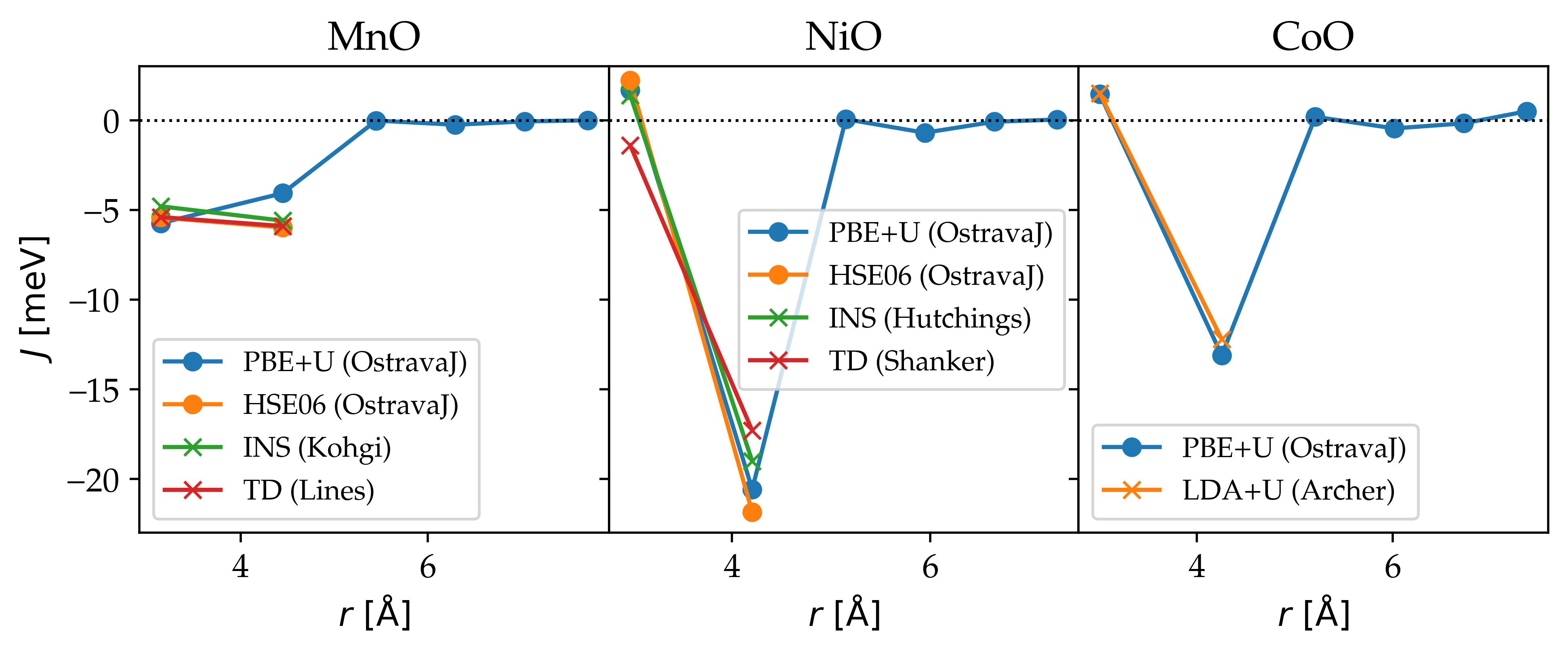}
    \caption{Magnetic exchange interactions of a) MnO: calculated with PBE+U and HSE06 functionals, from inelastic neutron scattering \cite{Kohgi}, from thermodynamic data \cite{Lines}; b) NiO: calculated with PBE+U and HSE06 functionals, from inelastic neutron scattering \cite{Hutchings}, from thermodynamic data \cite{Shanker}; and CoO: calculated with PBE+U approach, calculated with LDA+U in \cite{Archer}.}
    \label{fig:TMO-Js}
\end{figure}

\autoref{fig:TMO-Js} shows magnetic exchange interactions calculated with OstravaJ, compared to other computational or experimental works. Unlike in other works, we were able to calculate not only the first 2 nearest exchange interactions, but also 4 further longer-range interactions, and confirm that they are indeed very small. We can see a larger discrepancy in MnO between our calculation employing the PBE+U approach and the experimental figures. However, this can be attributed to an inaccuracy in the PBE functional. The match to experiment was significantly improved after using the HSE06 hybrid functional, which is much more accurate, albeit very computationally expensive.

\begin{table}[H]
\centering
\begin{tabular}{|c|c|c|c|}
\hline
 & MnO & NiO & CoO \\
 \hline
 OstravaJ - PBE+U - MFA & 83 K & 444 K & 268 K \\
 \hline
 expt. Js + MC & 85 K \cite{Kohgi} & 340 K \cite{Hutchings} & 210 K \cite{Archer} \\
 \hline
\end{tabular}
\caption{Calculated Néel temperatures for MnO, NiO and CoO using a) exchange interactions from this work and mean-field approximation, b) experimental exchange interactions with Monte Carlo calculations \label{tab:TMO-temp}}
\end{table}

\autoref{tab:TMO-temp} shows Néel temperatures calculated from the exchange interactions using mean-field theory, compared to other sources. For NiO and CoO, the mean-field temperature is significantly larger than the Monte Carlo one. This is most likely due to a well-known overestimation of critical temperatures by the mean-field approximation method.

\subsection{Metals}

The capability of OstravaJ to calculate long-range exchange interactions can be very well shown on a few important metallic materials, which do not have an insulating gap and, unlike in TMOs, the longer-range interactions are non-negligible. In this section, we calculate exchange interactions in hexagonal-closely-packed Cobalt (hcp Co), face-centered-cubic Cobalt (fcc Co), face-centered-cubic Nickel (fcc Ni) and body-centered-cubic Iron (bcc Fe). These materials have been chosen for the availability of published data for comparison.

For each system, we manually selected the exchange interaction cutoff and the size of the supercell such that OstravaJ was able to find a sufficient amount of independent configurations and the supercell was not too large, so that the calculations were not too computationally complex. The selection is shown in \autoref{tab:metal-cutoff}

\begin{table}[H]
    \centering
    \begin{tabular}{|c|c|c|c|c|}
    \hline
         & hcp Co & fcc Co & fcc Ni & bcc Fe \\
         \hline
         \makecell{no. of nearest\\exchange interactions} & 8 & 6 & 6 & 6 \\
         \hline
         distance cutoff [\AA] & 5.9 & 6.1 & 6.1 & 5.7 \\
         \hline
         supercell size & $4 \times 2 \times 2$ & $4 \times 4 \times 2$ & $4 \times 4 \times 2$ & $4 \times 4 \times 2$ \\
    \hline
    \end{tabular}
    \caption{Choice of exchange interaction cutoff and supercell size for calculations of metals.}
    \label{tab:metal-cutoff}
\end{table}

Total energies of all magnetic configurations were then calculated by VASP with the PBE exchange-correlation functional and parameters shown in \autoref{tab:metal-params}.

\begin{table}[H]
    \centering
    \begin{tabular}{|c|c|c|c|c|}
     \hline
     & \makecell{hcp Co} & \makecell{fcc Co} & \makecell{fcc Ni} & \makecell{bcc Fe} \\
     \hline
     lattice parameter [\AA] & 2.48 & 3.51 & 4.20 & 4.20\\
     \hline
     energy cutoff [eV] & 400 & 500 & 400  & 400 \\
     \hline
     k-grid & $8 \times 15 \times 8$ & $10 \times 10 \times 20$ & $8\times8 \times 16$ & $6 \times 6 \times 12$\\
     \hline
     \makecell{energy convergence\\criterion [eV]} & $10^{-7}$ & $10^{-6}$ & $10^{-6}$ & $10^{-7}$\\
     \hline
     \makecell{no. of valence electrons} & 9 & 17 & 10 & 8 \\
     \hline
\end{tabular}
    \caption{Important parameters for the DFT calculations of metals.}
    \label{tab:metal-params}
\end{table}

\begin{figure}[H]
    \centering
    \includegraphics[width=\linewidth]{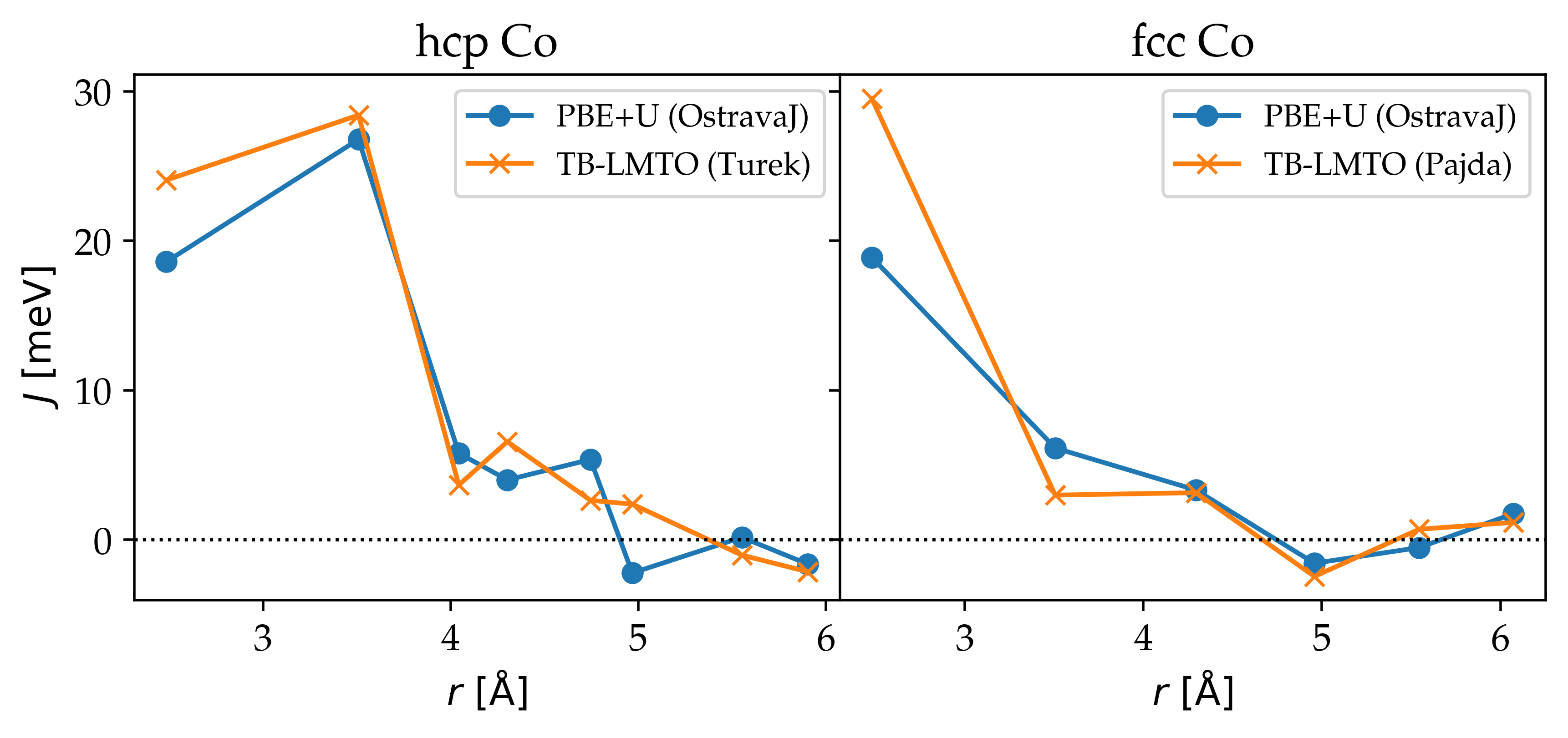}
    \caption{Exchange interactions for a) hcp Co: calculated using OstravaJ and PBE+U approach, via TB-LMTO method \cite{turek_ab_2003}, b) fcc Co: calculated using OstravaJ and PBE+U approach, via TB-LMTO method \cite{pajda_ab_2001}.}
    \label{fig:Co-Js}
\end{figure}

\begin{figure}[H]
    \centering
    \includegraphics[width=\linewidth]{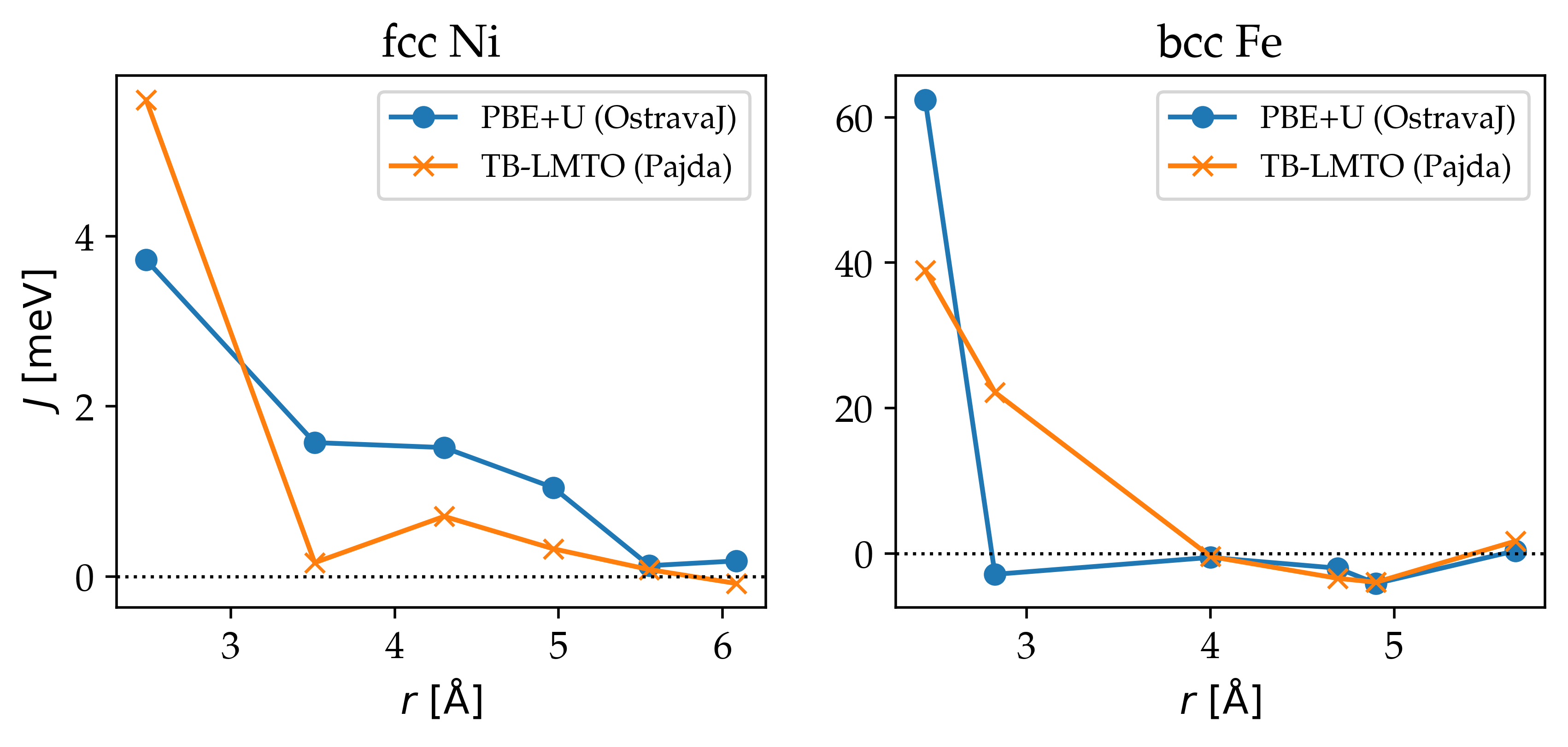}
    \caption{Exchange interactions for a) fcc Ni: calculated using OstravaJ and PBE+U approach, via TB-LMTO method \cite{pajda_ab_2001}, b) bcc Fe: calculated using OstravaJ and PBE+U approach, via TB-LMTO method \cite{pajda_ab_2001}.}
    \label{fig:NiFe-Js}
\end{figure}

The main results of the calculations, the exchange interaction energies, are shown in \autoref{fig:Co-Js} for hcp Co and fcc Co and \autoref{fig:NiFe-Js} for fcc Ni and bcc Fe. Energies calculated via the TB-LMTO (tight-binding linear-muffin-tin-orbital) method \cite{turek_ab_2003,pajda_ab_2001} are shown for comparison.

The results from this work and from the TB-LMTO method for the Cobalt systems are qualitatively in agreement -- the sign and the magnitude of the nearest interaction energies are comparable, and the energies diminish with distance in a similar fashion.

Results for fcc Ni and bcc Fe, unfortunately, show a significant deviation from the TB-LMTO method.

\begin{table}[H]
    \centering
    \begin{tabular}{|cc|c|c|}
    \hline
    & & $T_\mathrm{C}$ [K] & $D~\mathrm{[meV\cdot\AA^2]}$ \\
    \hline
    hcp Co& PBE+U & 1442 (MFA), 1281 (RPA)& 597 \\
    & TB-LMTO \cite{pajda_ab_2001} & 1673 (MFA) & --  \\
    \hline
    fcc Co& PBE+U & 1442 (MFA), 1281 (RPA) & 597 \\
    & TB-LMTO \cite{pajda_ab_2001}& 1645 (MFA), 1311 (RPA) & 663 \\
    & exp. & 1388 -- 1398 & 580 \cite{pauthenet_experimental_1982}, 510 \cite{shirane_spin_1968} \\
    \hline
    fcc Ni& PBE+U & 416 (MFA), 367 (RPA) & 597 \\
    & TB-LMTO \cite{pajda_ab_2001}& 397 (MFA), 350 (RPA)& 756  \\
    & exp. & 624--631 & 555 \cite{mook_temperature_1973}, 422 \cite{pauthenet_experimental_1982} \\

    \hline
    bcc Fe& PBE+U & 1530 (MFA), 708 (RPA) & 103  \\
    & TB-LMTO \cite{pajda_ab_2001} & 1414 (MFA), 950 (RPA) & 250 \\
    & exp. & 1044--1045 & 280 \cite{pauthenet_experimental_1982}, 330 \cite{shirane_spin_1968} \\
    \hline
    \end{tabular}
    \caption{Comparison of critical temperature and spin-wave stiffness between this work, TB-LMTO method \cite{pajda_ab_2001} and experimental results.}
    \label{tab:metal-props}
\end{table}

Exchange interaction energies are by themselves difficult to measure experimentally. However, they can be used to calculate critical temperature and spin-wave stiffness ($T_{\mathrm{C}}$, $D$) which are experimentally accessible quantities.  \autoref{tab:metal-props} shows the comparison of $T_{\mathrm{C}}$ and $D$ among values calculated from OstravaJ method, TB-LMTO method, and measured experimentally. The majority of the figures agree within tens of percent -- an adequate accuracy in the context of ab-initio methods. One exception is the spin-wave stiffness in bcc Fe -- the OstravaJ method reports roughly $D_\mathrm{OJ} \approx 100~\mathrm{meV \cdot \AA^2}$, while the experimental value is roughly $D_\mathrm{ex} \approx 300~\mathrm{meV \cdot \AA^2}$.

\section{Conclusion}

The total energy difference method is a widespread method for calculating exchange interaction energies for a wide variety of magnetic materials. However, a challenging part of this method is the search for and selection of suitable magnetic configurations. We propose a novel method of traversing the magnetic configuration space and overcoming this challenge.

We introduce the OstravaJ code, which automates most parts of the calculation via the total energy difference method, including the search for suitable magnetic configurations. It can thus be easily used in the field of high-throughput computational materials science.

Furthermore, the total energy difference method and the OstravaJ code were benchmarked on a set of several sample materials. We have shown that the total energy difference method is a viable computational method and gives reasonable results. However, the results are not in full agreement with calculations via the TB-LMTO method, which uses the LKAG formula. This can be attributed to the fact that this method is based on perturbing the magnetic ground state, while the total energy difference method also takes into account excited magnetic configurations. Thus, for materials which do not perfectly fit the Heisenberg model picture, the results are going to differ.

\section{Acknowledgments}

The authors acknowledge grant No. 22-35410K by Czech Science Foundation.
This work was supported by the Ministry of Education, Youth and Sports of the Czech Republic through the e-INFRA CZ (ID:90254).

 \bibliographystyle{elsarticle-num}
 \bibliography{cas-refs}





% \end{thebibliography}
\end{document}